\documentclass[11pt,a4paper]{article}
\usepackage{amssymb, amsmath,framed,mathrsfs}
\usepackage[colorlinks]{hyperref}
\newcommand{\rem}[1]{}
\newcommand{\de}{{\rm d}}

\newcommand{\bz}{{\mathbf{z}}}

\newcommand{\bzeta}{\boldsymbol{\zeta}}
\newcommand{\tbzeta}{\,\tilde{\!\boldsymbol{\zeta}}}

\newcommand{\bZeta}{{_{\,}\widehat{\!\boldsymbol{Z}}}}

\newcommand{\tW}{{\widetilde{W}}}

\newcommand{\mP}{\ensuremath{\langle \widehat{P} \rangle}}
\newcommand{\mQ}{\ensuremath{\langle \widehat{Q} \rangle}}
\newcommand{\mPs}{\ensuremath{\langle \widehat{P}^{2} \rangle}}
\newcommand{\mQs}{\ensuremath{\langle \widehat{Q}^{2} \rangle}}
\newcommand{\mQP}{\ensuremath{\langle \widehat{Q}\widehat{P} \rangle}}

\newcommand{\beq}{\begin{equation}}
\newcommand{\eeq}{\end{equation}}
\newcommand{\ben}{\begin{eqnarray}}
\newcommand{\een}{\end{eqnarray}}

\begin{document}

\title{Hamiltonian approach to Ehrenfest expectation values\\and Gaussian quantum states}
\author{Esther Bonet-Luz, Cesare Tronci\\
\it\footnotesize Department of Mathematics, University of Surrey, Guildford GU2 7XH, United Kingdom\\
}
\date{}


\maketitle

\begin{abstract}
The dynamics of quantum expectation values is considered in a geometric setting. First, expectation values of the  canonical observables are shown to be equivariant momentum maps for the action of the Heisenberg group on quantum states. Then, the  Hamiltonian structure of Ehrenfest's theorem is shown to be Lie-Poisson for a semidirect-product Lie group, named the \emph{Ehrenfest group}. {\color{black}The underlying Poisson structure produces classical and quantum mechanics as special limit cases.} In addition,  quantum dynamics is expressed in the frame of the expectation values, in which the latter undergo canonical Hamiltonian motion. In the case of Gaussian states, expectation values dynamics couples to second-order moments, which also enjoy a momentum map structure. Eventually, Gaussian states are shown to possess a Lie-Poisson structure associated to {\color{black}another semidirect-product group}, which is called the Jacobi group. This structure produces the energy-conserving variant of a class of Gaussian moment models previously appeared in the chemical physics literature.
\end{abstract}

\bigskip

%

\section{Introduction}

The expectation value dynamics
 for quantum canonical observables $\bZeta=(\boldsymbol{\widehat{Q}},\boldsymbol{\widehat{P}})$ 
 has always attracted much attention, especially in connection to the correspondence principle. In this context, quantum effects are produced by noncommutativity of the canonical observables, thereby leading to Heisenberg uncertainty. In turn, uncertainty is associated to distributional effects that are typically encoded in  Schr\"odinger's probability density $\left|\psi(\boldsymbol{x})\right|^2$. Then, one is led to look at statistical moments in order to obtain information about macroscopic quantities. 
Ehrenfest's equations for expectation value dynamics read
 \beq
i\hbar\frac{\de}{\de t}\langle{\bZeta}\rangle =\big\langle \big[ {\bZeta},\widehat{H} \big] \big\rangle
\,,
\label{Ehrenfest}
\eeq
where {\color{black}$\hat{H}$ is the quantum Hamiltonian operator and the average is computed with respect to the density matrix (that is, $\rho=\psi\psi^\dagger$ for pure states). Ehrenfest dynamics has recently been  considered from a geometric perspective \cite{CGMa} and the present paper uses similar geometric methods to identify its Hamiltonian structure.} 

{\color{black}The relation \eqref{Ehrenfest} can also be expressed in terms of the Wigner phase-space function ${W}(\boldsymbol{q},\boldsymbol{p})$ associated to $\rho$ \cite{Wigner}. This is conveniently written in Dirac notation, so that in one dimension the ket $|x\rangle$ is the eigenvector given by ${\widehat{Q}}|x\rangle=x|x\rangle$, while the bra vector is defined as $\langle x|:=|x\rangle^\dagger$. Upon generalizing to three dimensions, the Wigner function is associated to $\rho$ by the relations \cite{Zachos}
\begin{align*}
    {W}(\boldsymbol{q},\boldsymbol{p}) &= \frac{1}{(2\pi\hbar)^3} \int\! \left\langle \boldsymbol{q}+\frac{\boldsymbol{x}}2\right|\rho \left|\boldsymbol{q}-\frac{\boldsymbol{x}}2\right\rangle\, e^{-\frac{i\boldsymbol{p}\cdot\boldsymbol{x}}{\hbar}} \de^3 x \,,
\qquad
    \rho=\int \left|\boldsymbol{q}-\frac{\boldsymbol{x}}2\right\rangle e^{\frac{i\boldsymbol{p}\cdot\boldsymbol{x}}\hbar\,} W\left\langle \boldsymbol{q}+\frac{\boldsymbol{x}}2\right|\de^3x\,\de^3q\,\de^3p
\end{align*}
The mapping  that gives the Wigner function  from the density matrix (first relation above) is known as {\it Wigner transform of $\rho$}, while the inverse mapping (second relation above) is called {\it inverse Wigner transform}, or more often {\it Weyl transform of $W(\boldsymbol{q},\boldsymbol{p})$}. These mappings generate a one-to-one correspondence between operators and  phase-space functions.
The Wigner function is a quasi-probability distribution (i.e., it can be negative) and is sometimes simply called `Wigner distribution'. Its evolution reads   $\partial_t W=\left\{\!\left\{ H,W \right\}\!\right\}$, where ${H}(\boldsymbol{q},\boldsymbol{p})$ is  the Wigner transform of $\widehat{H}$ (also known as the \emph{Weyl symbol}) and $\left\{\!\left\{ \cdot,\cdot \right\}\!\right\}=\left\{ \cdot,\cdot \right\}_{c}+O(\hbar^2)$ is the Moyal bracket \cite{Moyal,Zachos} carrying the quantum noncommutative deviation (of order $O(\hbar^2)$) from the classical Poisson bracket  $\left\{ f,g \right\}_{c}=\partial_{\boldsymbol{x}}f\cdot\partial_{\boldsymbol{p}}g-\partial_{\boldsymbol{x}}g\cdot\partial_{\boldsymbol{p}}f$. The explicit definition of the Moyal bracket operator is rather involved and is omitted here. However, we shall recall that, although it does not identify a Poisson structure (as the Leibniz product rule is not satisfied), the Moyal bracket  defines a Lie bracket structure.
In this framework, Ehrenfest's theorem for the expectation values $\langle{\bzeta}\rangle=\int\!\bzeta\,W(\bzeta,t)\,\de^6\zeta$ is taken into the form
$
{\de}\langle{\bzeta}\rangle/{\de t} = \left\langle\left\{\!\left\{ {\bzeta},{H} \right\}\!\right\}\right\rangle
$, 
where $\bzeta=(\boldsymbol{q},\boldsymbol{p})$ is the phase-space coordinate.
At this stage, we recall another fundamental property of the Moyal bracket of two phase-space functions: 
 whenever one of these two is a second-degree polynomial, the Moyal bracket drops to the classical Poisson bracket.  {\color{black}Consequently, $\left\{\!\left\{ {\bzeta},{H} \right\}\!\right\}=\left\{ {\bzeta},{H} \right\}_c$ and the Wigner-Moyal formulation of Ehrenfest's theorem reads} 
\[
\frac{\de}{\de t}\langle{\bzeta}\rangle = \left\langle\left\{ {\bzeta},{H} \right\}_{c}\right\rangle,
\]
which coincides with classical expectation value dynamics (as it arises from the classical Liouville equation $\partial_t \varrho=\left\{ H,\varrho \right\}_c$), except for the fact that the averages are computed with respect to the quantum Wigner distribution $W$ (whose evolution accounts for quantum noncommutativity).

In the particular case when $\widehat{H}$ (or its Weyl symbol $H(\bzeta)$) is a quadratic polynomial, the Wigner equation coincides with the classical Liouville equation and it is solved by a Gaussian distribution on phase-space. In this case, an initial Gaussian evolves in time by changing its mean and variance, so that the latter moment tends to zero  in the formal limit $\hbar\to0$ (i.e. the Gaussian tends to a delta function or point particle).  See \cite{Littlejohn} for a geometric treatment of semiclassical Gaussian state evolution. Moreover, for quadratic Hamiltonians} the Ehrenfest theorem reduces to classical particle dynamics, that is 
 \[
\dot{\bz}=\Bbb{J}\nabla_{\!\bz} H(\bz)
\,,
\] 
where $\Bbb{J}$ is the canonical Poisson tensor $\Bbb{J}^{ij}=\{\zeta^i,\zeta^j\}_c$ and we have defined $\bz:=\langle{\bzeta}\rangle$. Then, one is led to conclude that the expectation values follow canonical Hamiltonian trajectories as long as an initial wavepacket keeps narrow in time, so that the expansion
\beq
H(\bzeta)\simeq H(\bz)+(\bzeta-\bz)\cdot\nabla H(\bz)+\frac12(\bzeta-\bz)\cdot\nabla \nabla H(\bz)(\bzeta-\bz)
\label{Taylor}
\eeq
is justified and  $\left\langle\left\{ {\bzeta},{H} \right\}_{c}\right\rangle\simeq\Bbb{J}\nabla H(\bz)$. However, this statement leads to interpretative questions \cite{BaYaZi} and  there is no reason why the approximation above should hold in the general case. 

Indeed, when $\widehat{H}$ is not quadratic, an initial Gaussian {\color{black}Wigner function} will evolve into different (non Gaussian) phase-space profiles, due to the effect of the noncommutative terms  in the Moyal bracket. 
The time by which these quantum corrections become important (so that the initial Gaussian profile spreads over macroscopic scales) is known as \emph{Ehrenfest time}  and is related to the possible breakdown of classical-quantum correspondence in quantum chaos \cite{KaZaZu02}. Then, a useful way to quantify the classical-quantum differences due to noncommutativity is through the difference between solutions of the classical Liouville and quantum Wigner equations (both initiated in the same  state), respectively. In recent years, this has been attempted by comparing the dynamics of quantum and classical statistical moments \cite{Ballentine, Brizuela}, {\color{black}that is the moments of the quantum Wigner function and the moments of the classical Liouville distribution, respectively}. Still, the relation between expectation value dynamics and canonical Hamiltonian motion remains unclear and the present paper aims to shed some light on this point. For example, in \cite{BoSh} {\color{black}(see equations (6.2) therein)} the expectation values $\bz=\langle{\bzeta}\rangle$ are shown to obey canonical Hamiltonian dynamics (even for non-Gaussian states) when the energy is expressed in terms of moments $\langle(\bzeta-\bz)^k\rangle$. {\color{black}However,  this result still lacks a more fundamental description in terms of quantum state dynamics.}

Without entering further the difficult questions concerning the interpretation of Ehrenfest's equations, this paper unfolds their geometric properties and  expresses the quantum evolution in the phase-space frame co-moving with the expectation values $\bz(t)$. More particularly, if one writes the total energy {\color{black}$h(W)=\int {W}(\bzeta,t)\,{H}(\bzeta)\,\de^6\zeta$ in terms of the Wigner function $W(\bzeta,t)$ and the expectation values $\bz$ (that is $h(W)=\tilde{h}(W,\bz)$), introducing the fluctuation variables $\tbzeta=\bzeta-\bz$ so that $W(\bzeta,t)=\widetilde{W}(\tbzeta,t)$, yields the Ehrenfest equations in canonical form (see Section \ref{Sec:changeframe})
\begin{equation}\label{newEhrenfest}
\dot{\bz}=\Bbb{J}\nabla_{\!\bz} {h}(\widetilde{W},\bz)
\,,
\end{equation}
where ${h}(\widetilde{W},\bz)=\tilde{h}(W,\bz)$ coincides with the total energy  of the quantum system}. As a new result, this paper also identifies the dynamics of the quantum state $\widetilde{W}(\tbzeta,t)$ accompanying the evolution of $\bz$. Interestingly enough, equation \eqref{newEhrenfest} is identical in form to the equation of motion of a classical particle in the Ehrenfest mean-field model of mixed classical-quantum dynamics \cite{Ehrenfest,Wyatt}. 
To avoid confusion, it is important to point out that \eqref{newEhrenfest} does not mean that the expectation values follow classical trajectories, as they would be obtained by the classical limit of a quantum system. Indeed, while the canonical structure implies  Hamiltonian trajectories of classical type, these trajectories do \emph{not} coincide with those of the classical physical system, which in turn would be obtained {\color{black}upon replacing ${h}(\widetilde{W},\bz)$ by the classical Hamiltonian function $\mathsf{h}(\bz)$.  This fact avoids the possibility of contradiction in equation \eqref{newEhrenfest}, which is to be coupled to the evolution of the relative quantum distribution $\widetilde{W}$.  We address the reader to \cite{ChMa} for further remarks on the geometric analogies and differences  between classical and quantum dynamics. 

In this paper, we shall focus on the geometric nature of the classical-quantum coupling that emerges from expectation value dynamics. The first goal   is to present (in Section \ref{sec:HamEhr}) a new formulation of Ehrenfest's theorem in terms of a classical-quantum Poisson bracket that couples the classical canonical bracket to the Poisson structure underlying quantum dynamics. This classical-quantum bracket naturally incorporates classical and quantum mechanics as special cases. However, as it was  briefly mentioned earlier, this work is not meant to provide a new interpretation of the classical-quantum correspondence. Rather, one of the targets of this paper  is to express (in Section \ref{Sec:changeframe}) the quantum dynamics in the phase-space frame co-moving with the expectation values, so that Ehrenfest equations possess a standard canonical form independently of the quantum state (not necessarily Gaussian). This canonical structure was found in \cite{BoSh} in the context of quantum cosmology: in this case, equation \eqref{newEhrenfest} is accompanied by the evolution of the moments $\langle{\tbzeta}$$^{\,k}\rangle=\langle(\bzeta-\bz)^k\rangle$. In the present paper, we replace the moment hierarchy by the  evolution of the quantum state $\widetilde{W}(\tbzeta,t)$ in the co-moving frame, where the entire Ehrenfest theorem is rewritten explicitly. 
The second part of the paper specializes this construction to consider the evolution of Gaussian states, as they are identified with Gaussian Wigner functions on phase-space. As a result, a Poisson bracket structure is provided in Section \ref{Gaussian} for the Gaussian moments $\langle\bzeta\rangle$ and  $\langle\bzeta\bzeta\rangle$. These quantities enjoy an (equivariant) momentum map structure that confers them a Lie-Poisson bracket \cite{Holm,HoScSt,MaRa}. Then, this bracket is analyzed in detail and related to the role of metaplectic transformations for the particular case of Gaussian wavepackets. Eventually, this construction is applied to provide energy-conserving Gaussian closure models. More particularly, the characterization of Gaussian state dynamics enables providing energy-conserving variants of previous Gaussian moment models \cite{Prezhdo2,PrezhdoPereverzev} in which the role of energy conservation has previously posed some issues \cite{PrezhdoPereverzev2}. As a general comment, we wish to emphasize that this paper does not deal with the convergence issues that may emerge in quantum mechanics. In particular, probability densities are assumed to decay sufficiently fast so that expectation values converge at all times. 
}

\section{Hamiltonian structure of Ehrenfest's theorem\label{sec:HamEhr}}

We start this Section with a mathematical result that lies at the basis of equation \eqref{newEhrenfest}: expectation values are momentum maps for the standard  representation of the Heisenberg group $\mathcal{H}(\Bbb{R}^{2n})\simeq\Bbb{R}^{2n}\times S^1$ on the space of wavefunctions. {\color{black}For completeness, we recall the multiplication rule for the Heisenberg group:
\[
(\boldsymbol{z}_1,\varphi_1)(\boldsymbol{z}_2,\varphi_2)=\left(\boldsymbol{z}_1+\boldsymbol{z}_2,\,{\varphi_1+\varphi_1-\frac12\,\boldsymbol{z}_1\cdot\Bbb{J}\boldsymbol{z}_2}\right)\,.
\]}A momentum map is, generally speaking, given by the generating function of the canonical transformation associated with a certain symmetry group $G$. For a symplectic vector space $(V,\omega)$ carrying a (simplectic) $G-$representation, the momentum map $J(z)$ is given by $2\langle J(z),\xi\rangle:=\omega(\xi_{V}(z),z)$,  for all $z\in V$ and  all $\xi\in \mathfrak{g}$. Here, $\mathfrak{g}$ denotes the Lie algebra of $G$, $\xi_V(z)$ denotes the infinitesimal generator of the $G-$representation and $\langle\cdot,\cdot\rangle$ is the natural duality pairing {\color{black}on $\mathfrak{g}^*\times\mathfrak{g}$ (here, $\mathfrak{g}^*$ is the dual vector space of $\mathfrak{g}$)}. While the reader is referred to \cite{Holm,HoScSt,MaRa} for further explanations, here we specialize this definition to the Hilbert space  $\mathscr{H}$ of wavefunctions, endowed with the standard symplectic form $\omega(\psi_1,\psi_2)=2\hbar\,{\rm Im}\langle\psi_1|\psi_2\rangle$ (here, $\langle \psi_1|\psi_2\rangle:=\int\! \bar{\psi}_1(\boldsymbol{x})\psi_2(\boldsymbol{x})\,\de ^3 x$). {\color{black}Indeed, the symplectic structure of the Hilbert space of wavefunctions is of paramount importance for the results in this paper.} A momentum map associated to a unitary $G-$representation on the  quantum state space  is the map $\mathbf{J}(\psi)\in\mathfrak{g}^*$  given by $\left<\mathbf{J}(\psi),\xi\right>:=\hbar\left<i\xi(\psi)|\psi\right>$, where $\xi(\psi)$ is the infinitesimal generator of the group action $\Phi_g(\psi)$ {\color{black}(with $g\in G$). When $G$ is the whole group of unitary transformations, the corresponding momentum map is given by $\mathbf{J}(\psi)=-i\hbar\psi\psi^\dagger$ \cite{ChMa}. } For the Heisenberg group of phase-space translations, the action is the usual displacement operator of coherent state theory \cite{Perelomov}. {\color{black}Upon denoting $\boldsymbol{z}=(\boldsymbol{q},\boldsymbol{p})$, this action is given by}
\begin{equation}\label{HWAction}
\left( \Phi _{( \boldsymbol{z} ,\varphi)}( \psi) \right) \!( \boldsymbol{x}):= e^{-i \frac\varphi\hbar} e^{-i \frac{\boldsymbol{p} \cdot \boldsymbol{q}}{2\hbar}}e^{i \frac{\boldsymbol{p} \cdot \boldsymbol{x}}\hbar} \psi (\boldsymbol{x}-\boldsymbol{q}) 
\,,
\end{equation}
{\color{black}so that the infinitesimal generator reads $\xi(\psi)=-i\hbar^{-1}(\phi+\bzeta\cdot\Bbb{J}\bZeta)\psi$ and we have identified $\xi=(\bzeta,\phi)\in \mathfrak{h}(\Bbb{R}^{2n})$ (here, $\mathfrak{h}(\Bbb{R}^{2n})\simeq\Bbb{R}^{2n+1}$ denotes the Heisenberg Lie algebra  so that the pairing  $\langle\cdot,\cdot\rangle$ on $\mathfrak{h}(\Bbb{R}^{2n})^*\times\mathfrak{h}(\Bbb{R}^{2n})$ is the standard dot product on $\Bbb{R}^{2n+1}$). Then, upon computing $\left<i\xi(\psi)|\psi\right>=\hbar^{-1}(\phi\|\psi\|^2+\bzeta\cdot\Bbb{J}\langle\bZeta\rangle)$,} we obtain the momentum map
\[
\mathbf{J}(\psi)=\left(\Bbb{J}\langle\bZeta\rangle,{\color{black}\|\psi\|^2}\right),
\]
so that, up to applying $\Bbb{J}$, the momentum map associated to the Heisenberg group representation identifies the expectation values of canonical observables. This momentum map is equivariant, that is $\operatorname{Ad}^*_{( \boldsymbol{z} ,\varphi)}\mathbf{J}(\psi)=\mathbf{J}(\Phi_{( \boldsymbol{z} ,\varphi)}(\psi))$ (here, $\operatorname{Ad}^*$ denotes the coadjoint representation for the Heisenberg group $\mathcal{H}(\Bbb{R}^{2n})$),
 and thus it is a Poisson map {\color{black}(see Section 12.4 in \cite{MaRa}).  Most importantly, this map takes the Poisson structure on the symplectic space  $\mathscr{H}$ of wavefunctions into the Lie-Poisson structure on $\mathfrak{h}(\Bbb{R}^{2n})^*$, i.e. the canonical Poisson bracket $\{\cdot,\cdot\}_c$ (up to a multiplicative normalization factor $\|\psi\|^2$). Notice that the momentum map $\mathbf{J}(\psi)$ induces an equivariant momentum map  $\mathbf{J}(\rho)=\big(\Bbb{J}\langle\rho|\bZeta\rangle,\operatorname{Tr}(\rho)\big)$  on the space of density matrices(here, $\rm Tr$ denotes the standard trace of trace-class operators): in this case, the infinitesimal action of the Heisenberg group reads $\xi(\rho)=-i\hbar^{-1}[\bzeta\cdot\Bbb{J}\bZeta,\rho]$. For convenience purposes, the remainder of this paper identifies quantum states with either density matrices or their Wigner transforms (unless otherwise specified). As we shall see, this leads to major simplifications in the treatment of Gaussian quantum states.}

The momentum map property of quantum expectation values suggests looking for the Poisson bracket structure of Ehrenfest's equations. We recall that the latter have to be accompanied by the evolution of the quantum state, this being given by a wavefunction, a density matrix or its Wigner function.  In order to find the Poisson bracket for the expectation values, we start with the following Poisson bracket \cite{Requist} for the quantum Liouville equation $i\hbar\dot\rho=[\widehat{H},\rho]$:
\beq
\{f,g\}{\color{black}(\rho)}=-i\hbar^{-1\!}
\left\langle \left[ \frac{\delta {f}}{\delta {\rho}},\frac{\delta {g}}{\delta {\rho}} \right] \right\rangle
,
\label{LiouvillePB}
\eeq
where  $[\cdot,\cdot]$ denotes the standard commutator {\color{black}and $f$ and $g$ are formally defined as function(al)s on the space of Hermitian operators.  
This Poisson bracket  returns the quantum Liouville equation as $\dot{\rho}=\{\rho,\delta h/\delta \rho\}$, with the Dirac Hamiltonian functional $h(\rho)=\langle \widehat{H}\rangle$ (total energy). Here, $\delta h/\delta \rho$ is a Hermitian operator and we identify  vector spaces of linear operators with their dual spaces by using the  pairing $\langle A,B\rangle={\rm Re}\langle A|B\rangle$, where $\langle A|B\rangle=\operatorname{Tr}(A^\dagger B)$ is the natural inner product. Since Ehrenfest dynamics advances both expectation values (denoted by $\bz=\langle\widehat{Z}\rangle$) and the quantum state (identified with the density matrix $\rho$), we allow to consider functionals of the type $\tilde{f}(\bz,\rho)$. In practice, this means that in \eqref{LiouvillePB} we allow for $f$ to depend on $\rho$ both explicitly and through the expectation value $\langle\widehat{Z}\rangle=\operatorname{Tr}(\rho\widehat{Z})$, so that $f(\rho)=\tilde{f}(\rho,\bz)$.}
{\color{black}Then, one can use the chain rule to write
\beq\label{chainrule}
 \frac{\delta}{\delta {\rho}}f(\rho)=\bZeta\cdot\frac{\delta}{\delta \bz}{\tilde{f}}(\bz,\rho)+ \frac{\delta}{\delta {\rho}}{\tilde{f}}(\bz,\rho)
\eeq
and replacing this in \eqref{LiouvillePB} yields
\beq
\{\tilde{f},\tilde{g}\}(\rho,\bz)=\{\tilde{f},\tilde{g}\}_{c}
+i\hbar^{-1\!}\left\langle
\left[\frac{\delta \tilde{g}}{\delta \bz}\cdot\bZeta,\frac{\delta \tilde{f}}{\delta \rho}\right]
-
\left[\frac{\delta \tilde{f}}{\delta \bz}\cdot\bZeta,\frac{\delta \tilde{g}}{\delta \rho}\right]
\right\rangle
-i\hbar^{-1\!}
\left\langle \left[ \frac{\delta \tilde{f}}{\delta {\rho}},\frac{\delta \tilde{g}}{\delta {\rho}} \right] \right\rangle
,
\label{hybridPB}
\eeq
where we have used the canonical commutation relations. The term $\{\tilde{f},\tilde{g}\}_{c}$ is multiplied by the normalization constant $\operatorname{Tr}(\rho)$ and in \eqref{hybridPB} we have set   ${\operatorname{Tr}(\rho)=1}$, which is the standard normalization of the density matrix. Without this particular choice, the change of variables $\bz\mapsto\Bbb{J}\bz$ and $\rho\mapsto -i\hbar\rho$ takes the Poisson bracket \eqref{hybridPB} into the (minus) Lie-Poisson \cite{Holm,HoScSt,MaRa} on the semidirect-product Lie algebra $\mathfrak{h}(\Bbb{R}^{2n})\,\circledS\,\mathfrak{u}(\mathscr{H})$ (here,  $\mathfrak{u}(\mathscr{H})$ is the Lie algebra of skew-Hermitian operators on  $\mathscr{H}$), endowed with the Lie bracket
\[
\Big[{(\bzeta_1,\phi_1,\xi_1)},(\bzeta_2,\phi_2,\xi_2)\Big]_{\!\mathfrak{h}(\Bbb{R}^{2n})\circledS\mathfrak{u}(\mathscr{H})\!}=\Big(\boldsymbol{0},-{\bzeta_1}\cdot\Bbb{J}\bzeta_2,\,i\hbar^{-1}\big[\bzeta_2\cdot\Bbb{J}\bZeta,\xi_1\big]-i\hbar^{-1}\big[\bzeta_1\cdot\Bbb{J}\bZeta,\xi_2\big] +\big[\xi_1,\xi_2\big]\Big).
\]
\rem{ 
The Lie-Poisson bracket is obtained as $\{f\,g\}_-=-\langle(\boldsymbol\nu,\alpha,\mu),\big[(\delta f/\delta\boldsymbol\nu,\delta f/\delta\alpha,\delta f/\delta\mu),(\delta g/\delta\boldsymbol\nu,\delta g/\delta\alpha,\delta g/\delta\mu)\big]\rangle$, with the natural pairing $\langle(\boldsymbol\nu,\alpha,\mu),(\bzeta,\phi,\xi)\rangle=\alpha\phi+\boldsymbol\nu\cdot\bzeta+{\rm Re}\operatorname{Tr}(\mu^\dagger\xi)$ (notice that $\alpha$ is a normalization constant, which is then set to 1).
} 
Here, we recall the general form of a ($\pm$)Lie-Poisson bracket, i.e.  $\{f,g\}_{\pm}(\mu)=\pm\langle\mu,[\delta f/\delta\mu,\delta g/\delta\mu]\rangle$ (where $\mu\in\mathfrak{g}^*$ is an element of some dual Lie algebra): the Lie-Poisson bracket \eqref{hybridPB} is obtained by choosing the minus sign and by specializing to the case $\mathfrak{g}=\mathfrak{h}(\Bbb{R}^{2n})\,\circledS\,\mathfrak{u}(\mathscr{H})$.
This semidirect-product Lie algebra  has been recently used in the variational formulation of the Ehrenfest theorem \cite{BlTr}, while it is shown here to emerge naturally from the momentum map structure of expectation values. Indeed, the occurrence of the canonical Poisson bracket is due to the fact that expectation values are Poisson momentum maps that take \eqref{LiouvillePB} into the classical canonical structure. We shall call the Lie algebra $\mathfrak{h}(\Bbb{R}^{2n})\,\circledS\,\mathfrak{u}(\mathscr{H})$ the \emph{Ehrenfest algebra} and its underlying Lie group $\mathcal{H}(\Bbb{R}^{2n})\,\circledS\,\mathcal{U}(\mathscr{H})$ the \emph{Ehrenfest group} (here, $\mathcal{U}(\mathscr{H})$ denotes the group of unitary operators on $\mathscr{H}$). The construction of the Ehrenfest group uses the group homomorphism provided by the representation \eqref{HWAction}, as it was presented in detail in Section IV.B of \cite{BlTr}. The bracket \eqref{hybridPB} will be called \emph{Ehrenfest bracket}: this is the first example of a classical-quantum bracket that couples the  canonical Poisson bracket underlying classical motion to the Lie-Poisson bracket \eqref{LiouvillePB} underlying quantum Liouville dynamics. However, notice that this Poisson bracket does not model the correlation effects  occurring in the interaction of quantum and classical particles. Indeed, Poisson bracket structures modeling the backreaction of a quantum particle on a classical particle have been sought for decades and are still unknown despite several efforts \cite{Aleksandrov,Anderson,CaSa,Kapral,Prezhdo,PrKi}. Rather, the Ehrenfest bracket governs the classical-quantum coupling (middle term in \eqref{hybridPB}) between  expectation value dynamics (first classical term in \eqref{hybridPB}) and quantum state evolution (last term in \eqref{hybridPB}) for the same physical system}.

{\color{black}At this point, in order to write down explicit equations of motion, one has to find the expression of the total energy $h(\rho)=\langle\widehat{H}\rangle$ in the form $h(\rho)=\tilde{h}(\rho,\bz)$. More particularly, we ask that $\tilde{h}(\rho,\bz)$ is linear in $\rho$, so that it can still be expressed as an expectation value. For example, the kinetic energy in one spatial dimension can be rewritten by using the relation $\langle\widehat{P}^2\rangle=p^2+\langle(\widehat{P}-p)^2\rangle$, with $p=\langle\widehat{P}\rangle$. In more generality, one can Taylor expand the original Hamiltonian operator (or simply parts of it) around the expectation values. This leads to an expression of the total energy $\langle\widehat{H}\rangle$ of the form 
 $\tilde{h}(\bz,\rho)=\langle \widehat{H}_\text{\tiny $CQ$}(\bz)\rangle$, where $\widehat{H}_\text{\tiny $CQ$}(\bz)=\delta \tilde{h}(\bz,\rho)/\delta \rho$ is a classical-quantum Hamiltonian operator that depends on the expectation values $\bz$ (and not explicitly on $\rho$). For example, in the simplest case of a free particle in one spatial dimension one writes $\widehat{H}_\text{\tiny $CQ$}(\bz)=(p^2/2)\boldsymbol{1}+(\widehat{P}-p\boldsymbol{1})^2/2$. Evidently,  analogous expressions hold for linear systems, such as the harmonic oscillator. Notice that $\widehat{H}_\text{\tiny $CQ$}(\bz)$ differs from the original quantum Hamiltonian $\widehat{H}$, although they generate (by definition) the same total energy $\langle\widehat{H}_\text{\tiny $CQ$}(\bz)\rangle=\langle\widehat{H}\rangle$.} Then, upon using the total energy expression $\tilde{h}=\langle \widehat{H}_\text{\tiny $CQ$}\rangle$, 
 the Poisson bracket \eqref{hybridPB} produces the equations
\begin{align}\label{michael}
&\dot{\bz} = \Bbb{J}\nabla_{\!\bz} \langle\widehat{H}_\text{\tiny $CQ$}\rangle -i\hbar^{-1} \big\langle \big[ {\bZeta},\widehat{H}_\text{\tiny $CQ$} \big] \big\rangle
\,,
\\
&i\hbar\dot{\rho} = \big[ \widehat{H}_\text{\tiny $CQ$}+ \nabla_{\!\bz} \langle\widehat{H}_\text{\tiny $CQ$}\rangle\cdot{\bZeta},\,\rho \big].
\label{miguel}
\end{align}
{\color{black}Here, it is important to emphasize that the relation $\bz=\langle\bZeta\rangle$ can be used only after evaluating the derivatives $\nabla_{\!\bz} \langle\widehat{H}_\text{\tiny $CQ$}\rangle$, which are to be computed by keeping $\bz$ and $\rho$ as independent variables (for example, one has $\nabla_{p}\big(p\langle\widehat{P}\rangle\big)=\langle\widehat{P}\rangle=p$).

Equations \eqref{michael} and \eqref{miguel}} were obtained in \cite{BlTr}, upon postulating a specific variational principle, based on analogies with the variational structure of the Ehrenfest mean-field model of mixed classical-quantum dynamics. Then, the action principle postulated in \cite{BlTr} is justified here in terms of its corresponding Hamiltonian structure. More importantly, we have shown how these equations are totally equivalent to the quantum Liouville equation. {\color{black}Indeed, these equations were derived from the Poisson bracket \eqref{LiouvillePB} for the density matrix evolution, without any sort of assumption or approximation: the only step involved was rewriting the total energy as $h(\rho)=\tilde{h}(\rho,\bz)$, which is no loss of generality as long as one term in the original Hamiltonian $\widehat{H}$ can be expanded around the expectation values.  Notice that, since equation \eqref{miguel} is indeed consistent with the ordinary Liouville equation, the classical-quantum Hamiltonian is constructed in such a way that it may differ from the original quantum Hamiltonian $\widehat{H}$ only by a phase factor, so that $\widehat{H}=\widehat{H}_\text{\tiny $CQ$}+ \nabla_{\!\bz} \langle\widehat{H}_\text{\tiny $CQ$}\rangle\cdot{\bZeta}+\phi\boldsymbol{1}$ (upon dropping the phase factor, this is simply a consequence of the chain rule relation \eqref{chainrule}): since by construction $\langle\widehat{H}_\text{\tiny $CQ$}\rangle=\langle\widehat{H}\rangle$, then $\phi=-\bz \cdot\nabla_{\!\bz} \langle\widehat{H}_\text{\tiny $CQ$}\rangle$ and the two Hamiltonians are related by
\[
\widehat{H}=\widehat{H}_\text{\tiny $CQ$}+({\bZeta}-\bz) \cdot\nabla_{\!\bz} \langle\widehat{H}_\text{\tiny $CQ$}\rangle
\,.
\]
Upon fixing a certain quantum system with Hamiltonian $\widehat{H}$, this is a consistency relation that is satisfied by the classical-quantum Hamiltonian. On the other hand, fixing a specific classical-quantum Hamiltonian, the above relation gives the corresponding quantum Hamiltonian operator.

Depending on the specific form of the  classical-quantum Hamiltonian}, equations (\ref{michael}) and (\ref{miguel}) allow for two different limits. Indeed, while Ehrenfest's theorem is obtained in the case $\nabla_{\!\bz} \langle\widehat{H}_\text{\tiny $CQ$}\rangle=0$ (that is, in the purely quantum case when the Hamiltonian operator is not written in terms of expectation values {\color{black}and $\widehat{H}_\text{\tiny $CQ$}=\widehat{H}$}), the phase-type operator $\widehat{H}_\text{\tiny $CQ$}=\mathsf{h}(\bz)\boldsymbol{1}$ yields another limit in which expectation values follow classical particle trajectories, i.e. $\dot{\bz} = \Bbb{J}\nabla \mathsf{h}(\bz)$. As noted in \cite{BlTr}, the latter case takes the quantum equation \eqref{miguel} into
\[
i\hbar\dot{\rho} =  \big[\nabla_{\!\bz} \mathsf{h}\cdot{\bZeta},\rho \big]
\,,
\]
which determines how the quantum state (together with Heisenberg's uncertainty) is carried  along the classical trajectories. {\color{black}The above Liouville equation is associated to a quantum Hamiltonian $\widehat{H}=\mathsf{h}(\bz)+({\bZeta}-\bz) \cdot\nabla\mathsf{h}(\bz)$ that is linear in $\bZeta$ and as such it generates coherent state evolution \cite{Perelomov} of an initial pure state $\rho_0=\psi_0\psi_0^\dagger$.} 
Notice that Wigner-transforming the above quantum equation of motion yields the classical Liouville dynamics 
\beq
\partial_t W(\bzeta)=\{\nabla_{\!\bz} \mathsf{h}\cdot{\bzeta},W(\bzeta)\}_c
\,,
\label{classicalcoherentstate}
\eeq 
so that noncommutative quantum effects are absent, as expected in coherent state dynamics. On one hand, this comes as no surprise, since coherent states are widely known to be classical states in quantum optics. (In quantum optics, the word `classical' does not refer to the motion of a physical classical particle, which would require a delta-like {\color{black}solution of \eqref{classicalcoherentstate} and is recovered only in the formal limit $\hbar\to0$ of Gaussian solutions of the type \eqref{gaussianwignerfunction}}). On the other hand, this yields a suggestive picture in which classical  trajectories carry a coherent quantum state, whose dynamics decouples from  classical motion.

While equations (\ref{michael}) and (\ref{miguel}) recover Ehrenfest's theorem as a special case, it is important to remark that they carry a redundancy, in the sense that equation (\ref{michael}) is simply the expectation value equation associated to (\ref{miguel}) (if the latter is interpreted as a nonlinear nonlocal equation). This redundancy is not new, since it already occurs in Ehrenfest's original equations. However, the redundancy of equation (\ref{michael}) can be eliminated by expressing the quantum dynamics in the frame of the expectation values. {\color{black}As we shall see, this operation separates the expectation values from the fluctuations arising from quantum uncertainty.}
While doing this is not generally trivial when using wavefunctions or density matrices, it becomes rather straightforward when using Wigner's phase-space description. We remark that  changes of frames in the configuration space for quantum dynamics were studied in the past \cite{Leblond,ScPl}, while changing the phase-space frame comes here as a new concept. This is the topic of the following Section.

\section{Quantum dynamics in the frame of expectation values\label{Sec:changeframe}}
As anticipated above, the splitting between quantum averages and fluctuations can be conveniently performed by writing the quantum dynamics in the frame of the expectation values. For this purpose, it is convenient to work in the Wigner phase-space formalism. Then, the equations (\ref{michael}) and (\ref{miguel}) become
\[
\dot{\bz}=\Bbb{J}\nabla_{\!\bz}\langle H_\text{\tiny $CQ$}\rangle+\langle \{\bzeta, H_\text{\tiny $CQ$}\}\rangle
\,,\qquad
\frac{\partial W}{\partial t}=\{\!\{H_\text{\tiny $CQ$},W\}\!\}+\nabla_{\!\bz}\langle H_\text{\tiny $CQ$}\rangle\cdot\{\bzeta,W\}
\,,
\]
where $H_\text{\tiny $CQ$}(\bz,\bzeta)$ is the Weyl symbol of the classical-quantum operator $\widehat{H}_\text{\tiny $CQ$}(\bz)$.
At this point, we perform the change of variables $\bzeta=\tbzeta+\bz$, so that the Weyl symbol of the Hamiltonian is rewritten as $H_\text{\tiny $CQ$}(\bz,\bzeta)=\widetilde{H}(\bz,\tbzeta)$ and analogously $W(\bzeta,t)=\widetilde{W}(\tbzeta,t)$. {\color{black}No confusion should arise here from the use of the tilde notation, which differs from that used in the previous Section.} Notice that this change of variables (or change of frame) preserves the canonical commutation relations and thus it does not affect Heisenberg's uncertainty principle. This change of variables 
\rem{ 
yields the  relations
\[
\nabla_{\!\bz} H=\nabla_{\!\bz} \widetilde{H}-\nabla_{\!\tbzeta\,}\widetilde{H}
\,,\qquad
\nabla_{\!\bzeta\,} H=\nabla_{\!\tbzeta\,}\widetilde{H}
\,,\qquad
\nabla_{\!\bzeta\,} W=\nabla_{\!\tbzeta\,}\widetilde{W}
\,,\qquad
\partial_t W=\partial_t \widetilde{W}-\dot{\bz}\cdot\nabla_{\!\tbzeta\,} \widetilde{W}
\]
Now, since 
$
\{\bzeta,H\}=\Bbb{J}\nabla_{\!\bzeta\,} H=\Bbb{J}\nabla_{\!\tbzeta\,}\widetilde{H}$,
the expectation value equation  $\dot{\bz}=\langle \Bbb{J}\nabla_{\!\bz} H\rangle+\langle \Bbb{J}\nabla_{\!\bzeta\,} H\rangle$ 
} 
returns \eqref{newEhrenfest}, 
which then holds {\color{black}regardless of the specific type of quantum state}. Also, the Wigner equation above transforms into
\rem{ 
\[
\partial_t \widetilde{W}+\nabla_{\!\bz}\langle \widetilde{H}\rangle\cdot\Bbb{J}\nabla_{\!\tbzeta\,} \widetilde{W}=\{\!\{\widetilde{H},\widetilde{W}\}\!\}+\langle \nabla_{\!\bz}\widetilde{H}-\nabla_{\!\tbzeta\,}\widetilde{H}\rangle\cdot\{\tbzeta,\widetilde{W}\}
\]
and thus
}  
$
\partial_t \widetilde{W}=\{\!\{\widetilde{H},\widetilde{W}\}\!\}-\langle\nabla_{\!\tbzeta\,}\widetilde{H}\rangle\cdot\{\tbzeta,\widetilde{W}\}
$.
Notice, the same result can be obtained in {\color{black}terms of Poisson structures} by changing the variables in the Ehrenfest Poisson bracket (\ref{hybridPB}), after expressing the latter in terms of Wigner functions. This can be done by considering the Lie-Poisson bracket for the Wigner equation \cite{BiBiMo} (recall the notation $\langle A(\bzeta)\rangle=\int\!W(\bzeta) A(\bzeta)\,\de\zeta$)
\beq\label{MoyalBracket}
\{f,g\}(W)=
\left\langle \left\{\!\!\left\{ \frac{\delta {f}}{\delta {W}},\frac{\delta {g}}{\delta {W}} \right\}\!\!\right\}\!\right\rangle
\,,
\eeq
and by following  similar steps as in the previous Section to obtain the Ehrenfest bracket (\ref{hybridPB}) in the form 
\beq\label{EhrenfestWigner}
\{f,g\}=\{f,g\}_{\bz}+\left\langle\left\{\frac{\delta {f}}{\delta \bz}\cdot\bzeta,\frac{\delta {g}}{\delta W}\right\}_{\!\!\bzeta}-\left\{\frac{\delta {g}}{\delta \bz}\cdot\bzeta,\frac{\delta {f}}{\delta W}\right\}_{\!\!\bzeta}\right\rangle
+
\left\langle\left\{\!\!\left\{\frac{\delta {f}}{\delta W},\frac{\delta {g}}{\delta W}\right\}\!\!\right\} \right\rangle
\,.
\eeq
Here, we have denoted $\{f,g\}_{\bz}=\nabla_{\!\bz} f\cdot\Bbb{J}\nabla_{\!\bz} g$ and analogously for $\{f,g\}_{\bzeta}$, while the Moyal bracket operates only on the phase-space quantum coordinates (so that $\left\{\!\left\{f,g\right\}\!\right\} =\left\{\!\left\{f,g\right\}\!\right\}_{\!\bzeta}$). {\color{black}Notice the abuse of notation, as $f$ and $g$ are very different in \eqref{MoyalBracket} and in \eqref{EhrenfestWigner} (compare with \eqref{LiouvillePB} and \eqref{hybridPB}). Then, changing variables to $\tbzeta=\bzeta-\bz$, so that $f(W,\bz)={\tt f}(\widetilde{W},\bz)$}, yields the relations

\[
\frac{\delta f}{\delta W}=\frac{\delta \mathtt{f}}{\delta \tW}
\,,\qquad\qquad
\left\{\!\!\left\{\frac{\delta {f}}{\delta W},\frac{\delta {g}}{\delta W}\right\}\!\!\right\}=
\left\{\!\!\left\{\frac{\delta \mathtt{f}}{\delta \tW},\frac{\delta \mathtt{g}}{\delta \tW}\right\}\!\!\right\}
\,,\qquad\qquad
\frac{\delta {f}}{\delta z}=\frac{\delta \mathtt{f}}{\delta z}-\int\!\tW\,\nabla_{\!\tbzeta\,}\frac{\delta \mathtt{f}}{\delta \tW}\,\de\tbzeta
\,,
\]
so that the Ehrenfest bracket becomes
\begin{align}\label{nonlocalbracket}
\{\mathtt{f},\mathtt{g}\}(\widetilde{W},\bz)=&\{\mathtt{f},\mathtt{g}\}_{\bz}+\left\langle\left\{\tbzeta,\frac{\delta \mathtt{f}}{\delta \tW}\right\}_{\!\tbzeta}\right\rangle\cdot\Bbb{J}\left\langle\left\{\tbzeta,\frac{\delta \mathtt{g}}{\delta \tW}\right\}_{\!\tbzeta}\right\rangle+
\left\langle\left\{\!\!\left\{\frac{\delta \mathtt{f}}{\delta \tW},\frac{\delta \mathtt{g}}{\delta \tW}\right\}\!\!\right\} \right\rangle,
\end{align}
where we used $\nabla_{\!\tbzeta\,}F=-\Bbb{J}\{\tbzeta,F\}_{\!\tbzeta\,}$ and the notation $\langle F(\tbzeta)\rangle:=\int\tW(\tbzeta)\,F(\tbzeta)\,\de^6\tilde{\zeta}$. Here, a useful feature of the Moyal bracket is the permutation property $
\int \!a(\bzeta)\{\!\{b(\bzeta),c(\bzeta)\}\!\}\,\de^6\zeta=\int\! c(\bzeta)\{\!\{a(\bzeta),b(\bzeta)\}\!\}\,\de^6\zeta
$
for any three functions $a,b,c\in C^\infty(\Bbb{R}^6)$.
Eventually, the total energy expression $\mathtt{h}(\bz,W)=\langle{H}_\text{\tiny$CQ$}(\bz,\bzeta)\rangle=\langle\widetilde{H}(\bz,\tbzeta)\rangle=\mathtt{h}(\bz,\widetilde{W})$ yields the coupled equations
\beq
\dot{\bz}=\{\bz, \langle\widetilde{H}\rangle\}_{\bz}
\,,\qquad\qquad
\frac{\partial \widetilde{W}}{\partial t} =
\big\{\!\big\{\widetilde{H},\widetilde{W}\big\}\!\big\}+\big\{\tbzeta\cdot\Bbb{J}\langle\{\tbzeta,\widetilde{H}\}_{\!\tbzeta}\rangle,\widetilde{W}\big\}_{\!\tbzeta}.
\label{neweqs}
\eeq
{\color{black}Here, $\langle\tbzeta\rangle=0$ is verified to be preserved in time as the zero-level set of the expectation value momentum map (see \cite{Tronci} for an analogue of this in classical moment dynamics).  Then, the redundancy occurring in \eqref{michael} and \eqref{miguel} has now been eliminated, since the expectation value equation is no longer computed from the  quantum Wigner-Moyal equation, which actually encodes the quantum deviations from the mean  variables $\bz$. 
\rem{ 
For completeness, we also rewrite the second equation above as a Schr\"odinger equation in the frame of expectation values: this reads $i\hbar\dot{\widetilde\psi} = \big( \widetilde{H}(\bz)-  \langle [\widetilde{Z}, \widetilde{H}(\bz)] \rangle\cdot\Bbb{J}\widetilde{Z} \big)\widetilde\psi$, where $\widetilde{H}(\bz)$ is the Hermitian operator (up to a phase possibly depending on $\bz$) obtained as the Weyl transform 
\[
\widetilde{H}(\bz)=2\int\!\de^3 \tilde{x}\,\de^3\tilde{y}\int\! \de^3\tilde{p}\, |\tilde{\mathbf{x}}+\tilde{\mathbf{y}}\rangle \widetilde{H}(\bz,\tbzeta) \langle\tilde{\mathbf{x}}-\tilde{\mathbf{y}}|
\]
(that is, the inverse Wigner transform \cite{Zachos}) of the symbol $\widetilde{H}(\bz,\tbzeta)$. 
}  
On the other hand, as already found in \cite{BoSh}, the first equation in \eqref{neweqs}} implies that expectation values evolve along canonical Hamiltonian trajectories with Hamiltonian $\langle\widetilde{H}\rangle$. This expectation value dynamics is different from classical Hamiltonian motion (one of the limits treated in the previous Section), as it carries higher moments of the type $\langle\tbzeta\tbzeta\dots\tbzeta\rangle$ (upon assuming that $\widetilde{H}$ is analytic) that are responsible for quantum deviations from the classical physical   trajectory \cite{BoSh}. {\color{black}The latter still emerges} when $\widetilde{H}(\bz,\tbzeta)$ is linear in the deviation coordinate $\tbzeta$, which is the case of coherent states. In this case, the relative distribution does not evolve in time and the relation $\partial_t\tW=0$ is simply equation \eqref{classicalcoherentstate} rewritten in the frame of expectation values.

Once quantum dynamics is expressed in the frame of expectation values, one can continue to compute moments $\langle\tbzeta\tbzeta\dots\tbzeta\rangle$ of the deviation coordinate. The identification of the Hamiltionian structure for these moments was carried out in \cite{BoSh}, in the context of quantum cosmology \cite{Bojowald}. This Hamiltonian structure was recently compared \cite{Brizuela} to that of classical moments of the Liouville equation  (see also \cite{HoLySc} for the classical case). The main difference between the results in \cite{BoSh} and those in the present paper is that here the moment hierarchy is replaced by the explicit evolution of the quantum state in the phase-space frame moving with $\bz(t)$. For example, in terms of the density matrix $\tilde{\rho}$ (as it is obtained by the Weyl transform of $\tW(\tbzeta,t)$), equations \eqref{neweqs} are replaced  by
\beq
\dot{\bz}=\{\bz, \langle\widetilde{H}_\text{\tiny ${CQ}$}\rangle\}_{\bz}
\,,\qquad\qquad
i\hbar\frac{\de \tilde{\rho}}{\de t}=\big[
\widetilde{H}_\text{\tiny ${CQ}$},\tilde{\rho}\big]+\frac{i}{\hbar}\big\langle\big[\widetilde{H}_\text{\tiny ${CQ}$},\widetilde{\boldsymbol{Z}}\big]\big\rangle\cdot\big[\Bbb{J}\widetilde{\boldsymbol{Z}}\,,\tilde{\rho}\big]
\label{neweqs2}
\eeq
where $\widetilde{\boldsymbol{Z}}=\bZeta-\bz$ and $\widetilde{H}_\text{\tiny ${CQ}$}(\bz)$ is the operator obtained as the Weyl transform of $\widetilde{H}(\bz,\tbzeta)$ (with respect to $\tbzeta$). This description has the advantage of overcoming the moment truncation problem, while it requires finding solutions of a nonlocal nonlinear quantum equation.

Notice that, unlike the moment approach, the present method does not require the Weyl symbol of the original Hamiltonian to be analytic. As an illustrative example in one dimension, here we consider a unit mass subject to a step potential, so that the phase-space Hamiltonian is $H(\bzeta)=\text{\sf\itshape p}^{\,2}/2+\mu\,\Theta (\text{\sf\itshape q})$, where $\mu$ is a physical parameter, $\Theta$ denotes Heaviside's step function, and we have used the notation  $\bzeta=(\text{\sf\itshape q},\text{\sf\itshape p})$.  Upon introducing the deviation coordinate $\tbzeta=\bzeta-\bz(t)$, one writes $H(\bz+\tbzeta)=({p}+\,\tilde{\!\text{\sf\itshape p}})^2/2+\mu\,\Theta({q}+\,\tilde{\!\text{\sf\itshape q}})$ with the notation $\bz(t)=(q(t),p(t))$.  Then, the total energy $\mathtt{h}=\langle H(\bz+\tbzeta)\rangle$ reads
\begin{align*}
\mathtt{h}(\bz,\widetilde{W})
&=\frac12\left(p^2+\langle\,\tilde{\!\text{\sf\itshape p}}^{\,2}\rangle\right)+\mu\langle\Theta({q}+\,\tilde{\!\text{\sf\itshape q}})\rangle
\\
&=\frac12\,p^2+\frac12\iint\tilde{\!\text{\sf\itshape p}}^{\,2}\,\widetilde{W}(\,\tilde{\!\text{\sf\itshape q}},\,\tilde{\!\text{\sf\itshape p}})\,\de\,\tilde{\!\text{\sf\itshape q}}\,\de\,\tilde{\!\text{\sf\itshape p}}+\mu\int_{-q}^{+\infty}\!\de\,\tilde{\!\text{\sf\itshape q}}\int\!\de\,\tilde{\!\text{\sf\itshape p}}\;\widetilde{W}(\,\tilde{\!\text{\sf\itshape q}},\,\tilde{\!\text{\sf\itshape p}})
\end{align*}
and thus $\delta\mathtt{h}/\delta\widetilde{W}=({p}^2+\,\tilde{\!\text{\sf\itshape p}}^{\,2})/2+\mu\,\Theta({q}+\,\tilde{\!\text{\sf\itshape q}})=\widetilde{H}(\bz,\tbzeta)$ is the effective Hamiltonian such that $\mathtt{h}=\langle H(\bz+\tbzeta)\rangle=\langle\widetilde{H}\rangle$. In this case, the function $\widetilde{H}(\bz,\tbzeta)$  is not analytic and the total energy $\langle\widetilde{H}\rangle$ cannot be expressed exactly in terms of the hierarchy of moments $\langle\tbzeta\tbzeta\dots\tbzeta\rangle$. However, in the present approach, one replaces moments by the explicit quantum evolution in the frame moving with the expectation values $\bz(t)$. In this framework, the latter evolve according to the first equation in \eqref{neweqs}, that is
\[
\dot{q}=p
\,,\qquad
\dot{p}=-\mu\int\!\widetilde{W}(-q,\,\tilde{\!\text{\sf\itshape p}})\,\de\,\tilde{\!\text{\sf\itshape p}}
\,.
\]
It is not surprising that the expectation values evolve under the time-dependent effective potential $\mu\int_{-q}^{+\infty}\!\de\,\tilde{\!\text{\sf\itshape q}}\int\!\de\,\tilde{\!\text{\sf\itshape p}}\;\widetilde{W}(\,\tilde{\!\text{\sf\itshape q}},\,\tilde{\!\text{\sf\itshape p}})$, so that  the step potential is modulated to allow for quantum tunneling. Notice that the term $\iint\tilde{\!\text{\sf\itshape p}}^{\,2}/2\,\widetilde{W}(\,\tilde{\!\text{\sf\itshape q}},\,\tilde{\!\text{\sf\itshape p}})\,\de\,\tilde{\!\text{\sf\itshape q}}\,\de\,\tilde{\!\text{\sf\itshape p}}$ in the total energy $\langle\widetilde{H}\rangle$ does not contribute to expectation value dynamics. In addition, the quantum state evolution is given by the Wigner equation \eqref{neweqs} for $\widetilde{W}(\,\tilde{\!\text{\sf\itshape q}},\,\tilde{\!\text{\sf\itshape p}},t)$. This is given by the following integro-differential equation
\[
\frac{\partial\widetilde{W}}{\partial t}-\left\{\!\!\left\{  \frac12\,\tilde{\!\text{\sf\itshape p}}^{\,2}+\mu\,\Theta({q}+\,\tilde{\!\text{\sf\itshape p}})  \,,\widetilde{W}\right\}\!\!\right\}
=-\mu\,\frac{\partial\widetilde{W}}{\partial \,\tilde{\!\text{\sf\itshape p}}}\int\!\widetilde{W}(-q,\,\tilde{\!\text{\sf\itshape p}})\,\de\,\tilde{\!\text{\sf\itshape p}}
\,,
\]
where a new nonlocal term emerging from the change of frame appears on the right hand side.  If this term is set to zero, one obtains the ordinary quantum dynamics in phase space for a particle subject to a potential step in $-q$. While explicit solutions can be constructed by taking the Wigner transform of the usual wavefunction of a unit mass subject to a step potential, the detailed study of the nonlinear nonlocal evolution of $\widetilde{W}$ is note among the purposes of this paper.

Notice, the whole treatment proceeds analogously for classical Liouville dynamics upon replacing Moyal brackets with Poisson brackets. Once more, this means that the essential difference between classical  and quantum statistical effects lies in the noncommutative terms (higher order in $\hbar$)  of the Moyal bracket. It is useful to emphasize that the conserved total energy $\langle\widetilde{H}\rangle$ does not depend explicitly on time, as it is readily seen by expanding $0=\de\langle\widetilde{H}\rangle/\de t=\partial_t\langle\widetilde{H}\rangle+\dot{\bz}\cdot\nabla_{\!\bz}\langle\widetilde{H}\rangle$ and using the first of \eqref{neweqs}, so that $\dot{\bz}\cdot\nabla_{\!\bz}\langle\widetilde{H}\rangle=0$. Notice that the relation $\partial_t\langle\widetilde{H}\rangle=0$ can also be verified explicitly by computing $\partial_t\langle\widetilde{H}\rangle=\int\!\widetilde{H}(\bz,\tbzeta)\,\partial_t\widetilde{W}(\tbzeta)\,\de^6\tilde{\zeta}=0$. Moreover, another consequence of the canonical nature of expectation value dynamics is the invariance of Poincar\'e's relative integral invariant:
$
\oint_{\gamma}\Bbb{J}\bz\cdot\de\bz=\oint_\gamma \langle\boldsymbol{\widehat{P}}\rangle\cdot\de\langle\boldsymbol{\widehat{Q}}\rangle=const.
$, 
for any loop $\gamma$ in phase-space moving with the Hamiltonian vector field $\{\bz,\langle\widetilde{H}\rangle\}_{\bz}.$

In the next Section, we focus on Gaussian quantum states, thereby restricting to consider only second-order moments. More particularly, we shall characterize the Hamiltonian structure of Gaussian state dynamics in terms of expectation values $\bz$ and covariance matrix  $\langle\tbzeta\tbzeta\rangle$.

\section{Hamiltonian structure of Gaussian quantum states\label{Gaussian}}

Once the dynamics of Ehrenfest expectation values has been completely characterized {\color{black}in terms of Poisson brackets}, one may consider higher order moments. The moment hierarchy does not close in the general case, although it is well known that it does for quadratic Hamiltonians. In the latter case, the moment algebra acquires an interesting structure, which is the subject of the present Section. Before entering this matter, we emphasize that quadratic Hamiltonians restrict to consider linear oscillator motion and so they are uninteresting for practical purposes. An interesting situation, however, occurs when the total energy $\langle H \rangle$ is computed with respect to a Gaussian state, so that higher moments are expressed in terms of the first two. This is the Gaussian moment closure for nonlinear quantum Hamiltonians.

Gaussian quantum states (a.k.a. squeezed states in quantum optics) have been widely studied over the decades in many different contexts, mostly quantum optics and physical chemistry. Recently, applications of Gaussian states in quantum information have also been proposed (see e.g. \cite{AdRaLe}).
Generally speaking, a Gaussian quantum state is identified with a Gaussian Wigner function of the form
\beq
G(\bzeta,t)=\frac{N}{\sqrt{\det\Sigma(t)}}\exp\!\left(-\frac1{2\hbar}\,(\bzeta-\bz(t))\cdot\Sigma(t)^{-1}(\bzeta-\bz(t))\right),
\label{gaussianwignerfunction}
\eeq
where $N$ is a normalising factor, $\bz=\langle\bzeta\rangle$ is the mean and $\hbar\Sigma=\hbar(\langle\bzeta\bzeta\rangle-\langle\bzeta\rangle\langle\bzeta\rangle)$ is the covariance matrix. The question of which covariance matrices are associated to genuine Gaussian quantum states was addressed in \cite{Littlejohn} for wavepackets and in \cite{Gracia,SiSuMu1} for more general mixed states. For example, when the Gaussian is too narrow, then  the corresponding density matrix is not positive-definite and  the uncertainty principle is violated. However, when the initial covariance matrix satisfies Heisenberg's principle (so that \eqref{gaussianwignerfunction} identifies a  quantum state), the latter holds at all times as the quantum uncertainty is preserved by the Wigner-Moyal equation $\partial_t G=\left\{\!\left\{ H,G \right\}\!\right\}$ for Gaussian states. {\color{black} We emphasize that the expression \eqref{gaussianwignerfunction} incorporates  the Wigner transform of Gaussian wavepackets \cite{Heller1,Heller2} as a special case.} 

If the linear form $\boldsymbol{w}$ and the quadratic form $S$ defining the quadratic Hamiltonian $H=\bzeta\cdot S\bzeta+\boldsymbol{w}\cdot\bzeta$ are  functions (possibly nonlinear) of the first and second order moments $\langle\bzeta\rangle$ and $\langle\bzeta\bzeta\rangle$, then a Gaussian initial state will remain a Gaussian under time evolution by changing its mean and variance. In more generality, a nonlinear (analytic) Hamiltonian will produce a total energy $\langle H\rangle =\int \!G(\bzeta) H(\bzeta)\,\de\zeta$ that can be expressed entirely in terms of first and second order moments, i.e. 
$$\langle H\rangle = h(\langle\bzeta\rangle,\langle\bzeta\bzeta\rangle)\,.$$
At this point, it is useful to restrict the Ehrenfest bracket \eqref{hybridPB} (or its phase space variant \eqref{EhrenfestWigner}) to functionals of Gaussian Wigner functions $G(\bzeta)$, depending only on the first two moments. For convenience, we shall denote $X=\langle\bzeta\bzeta\rangle/2$, so that $h(\bz,G)=\mathsf{H}(\bz,X)$. The corresponding Poisson structure is easily found by using the chain rule relation
\begin{align}
\frac{\delta f}{\delta G} =  \bz\cdot\!\left(\frac{\delta \mathsf{F}}{\delta \bz}  + \frac{1}{2}\frac{\delta \mathsf{F}}{\delta X}\bz\right), \label{EQ:chainr2}
\end{align}
in \eqref{MoyalBracket}. As a result, one finds 
\begin{align}
\lbrace \mathsf{F},\mathsf{G} \rbrace (\bz,X)= 
\{\mathsf{F},\mathsf{G}\}_{c}
+ \bz \cdot \left( \frac{\delta \mathsf{F}}{\delta X} \Bbb{J}\frac{\delta \mathsf{G}}{\delta \bz} - \frac{\delta \mathsf{G}}{\delta X} \Bbb{J} \frac{\delta \mathsf{F}}{\delta \bz} \right)
+ \text{Tr}\!\left( X\! \left[ \frac{\delta \mathsf{F}}{\delta X}\Bbb{J} \frac{\delta \mathsf{G}}{\delta X}-\frac{\delta \mathsf{G}}{\delta X}\Bbb{J} \frac{\delta \mathsf{F}}{\delta X} \right] \right) ,
\label{EQ:QmomPB}
\end{align}
along with the following equations of motion for an arbitrary total energy $\mathsf{H}(\bz,X)$:
\begin{align}\label{Jack1}
\dot{\bz} &= \{\bz,\mathsf{H}\}_c+\Bbb{J}\frac{\delta \mathsf{H}}{\delta X} \bz\,,
\\\label{Jack2}
\dot{X} &=  \left( \Bbb{J}\frac{\delta \mathsf{H}}{\delta X}X + X\frac{\delta \mathsf{H}}{\delta X}\Bbb{J} \right)
+ \frac{1}{2}\left( \Bbb{J}\left( \frac{\delta \mathsf{H}}{\delta \bz}\bz \right) + \left(\bz\frac{\delta \mathsf{H}}{\delta \bz}\right) \Bbb{J} \right).
\end{align}

The Poisson bracket \eqref{EQ:QmomPB} has appeared earlier in the literature \cite{GBTr,HoTr} in the context of classical Liouville (Vlasov) equations. This is no surprise, as second order moments are associated to quadratic phase-space polynomials, which do not involve the higher-order noncommutative terms in the Moyal bracket (Gaussian quantum states undergo classical Liouville-type evolution). Under the change of variables $X\mapsto X\Bbb{J}$ and $\bz\mapsto\Bbb{J}\bz$, the above moment bracket was shown  \cite{GBTr} to be Lie-Poisson on the Jacobi group 
\[
\operatorname{Jac}(\Bbb{R}^{2n})=\operatorname{Sp}(\Bbb{R}^{2n})\,\circledS\,\mathcal{H}(\Bbb{R}^{2n})\,,
\]
i.e. the semidirect product of the symplectic group with the Heisenberg group. Here, the semidirect-product structure is defined by the following action of the symplectic group on the Heisenberg group: $\Phi_S(\boldsymbol{z},\varphi)=(S\boldsymbol{z},\varphi)$, where $S\in \operatorname{Sp}(\Bbb{R}^{2n})$. Therefore, we conclude that the moments of any Gaussian quantum state evolve on coadjoint orbits of the Jacobi group. In particular, this means that the symplectic forms recently found to underlie the dynamics of Gaussian wavepackets \cite{Ohsawa,OhLe} are symplectic forms on coadjoint orbits of $\operatorname{Jac}(\Bbb{R}^{2n})$, which are determined by the usual Casimir invariants \cite{DoMa,HoLySc}
\[
C_j(\bz,X)=\frac1{2j}\operatorname{Tr\!}\left[\left(X-\frac12\bz\bz\right)\!\Bbb{J}\right]^{2j}
\,,
\]
where $j=1,2,3$ and $\operatorname{Tr}$ denotes the matrix trace. For example, setting $j=1$  and expanding yields (up to multiplicative factors) $C_1(\bz,X)=\det(2X-\bz\bz)$.

These results come as no surprise. The relation between the Jacobi group and Gaussian states has been known for decades in the theory of coherent states \cite{Berceanu2007b,Berceanu2009}, under the statement that wavepackets evolve under the action of the semidirect product $\operatorname{Mp}(\Bbb{R}^{2n})\,\circledS\,\mathcal{H}(\Bbb{R}^{2n})$ \cite{deGosson,Littlejohn}, where $\operatorname{Mp}(\Bbb{R}^{2n})$ is the metaplectic group (that is, the double covering of $\operatorname{Sp}(\Bbb{R}^{2n})$). Since the Lie algebra $\mathfrak{mp}(\Bbb{R}^{2n})$ of the metapletic group is isomorphic to that of the symplectic group (denoted by $\mathfrak{sp}(\Bbb{R}^{2n})$), then $\mathfrak{jac}(\Bbb{R}^{2n})\simeq\mathfrak{mp}(\Bbb{R}^{2n})\,\circledS\,\mathfrak{h}(\Bbb{R}^{2n})$.
In the present treatment, the symplectic group replaces the metaplectic transformations {\color{black}because  we have identified quantum states with  Wigner functions (which can account for pure as well as mixed states) rather than  wavefunctions. Indeed, while the symplectic group does not possess a representation on wavefunctions, it does possess a natural action on the space of Wigner (phase-space) functions as in classical mechanics and one may avoid dealing with the metaplectic representation \cite{Littlejohn}.
 This leads to an action of the Jacobi group on the space of  Wigner functions, which is given by 
\beq\label{JacobiAction}
(\Phi_{(\mathcal{S},\boldsymbol{z},\varphi)}(W))(\bzeta)
=W(\mathcal{S}\bzeta+\boldsymbol{z})
\,,
\eeq 
where $\mathcal{S}$ is a symplectic matrix and $(\boldsymbol{z},\varphi)$ is an element of the Heisenberg group. This action is given by the pullback of the Wigner function $W(\bzeta)$ by the phase-space transformation $\bzeta\mapsto \mathcal{S}\bzeta+\boldsymbol{z}$.

The emergence of a Lie-Poisson bracket for the first and second moments is also not surprising. Indeed, this is due to the fact that the moment triple $(\langle1\rangle,\bz,X)$ is itself an equivariant momentum map for the action \eqref{JacobiAction}; see Section III.C of \cite{GBTr}. (Here, we have formally denoted $\langle1\rangle=\int\!W(\bzeta)\,\de\zeta$). Then, Gaussian Wigner functions are identified with the moment couple $(\bz,X)$ and this identification enables the description of Gaussian state dynamics in terms of coadjoint orbits.
}

\rem{ 
infinitesimal action of the Jacobi algebra on the Hilbert space $\mathscr{H}$ of wavefunctions. {(We recall )} This infinitesimal action is  given by $\xi(\psi)=-i\hbar^{-1}(\phi+\bzeta\cdot\Bbb{J}\bZeta+\bZeta\cdot \Bbb{J}\tilde{S}\bZeta/2) \psi$, where we have identified $\xi=(\bzeta,\phi,\tilde{S})\in \mathfrak{jac}(\Bbb{R}^{2n})$ and $\tilde{S}\in \mathfrak{sp}(\Bbb{R}^{2n})$. The momentum map is then given by
\[
\mathbf{J}(\psi)=\left(\Bbb{J}\langle\bZeta\rangle,\langle
1\rangle,\frac12\langle\bZeta\bZeta\rangle\Bbb{J}\right),
\]
which identifies the first moments (upon suitably multiplying by $\Bbb{J}$).
Although the infinitesimal action producing this momentum map does not lift to a representation of the Jacobi group on $\mathscr{H}$, it does lift to an action of $\operatorname{Mp}(\Bbb{R}^{2n})\,\circledS\,\mathcal{H}(\Bbb{R}^{2n})$ \cite{Littlejohn} and this is how the metaplectic group comes into play. However, if one gives up on pure quantum states (wavefunctions) and allows for mixed states, then the Jacobi group has a natural action on the space of Wigner (phase-space) functions as in classical mechanics and one may avoid dealing with the metaplectic representation. Then, one also verifies that $\mathbf{J}$ is equivariant and thus it is a Poisson map \cite{MaRa,HoScSt}.
}   

\section{Gaussian moment models and energy conservation}
Notice that one can rewrite the above dynamics in terms of the covariance matrix $\Sigma\color{black}=2X-\bz\bz$.
 This is easily done by restricting the bracket \eqref{nonlocalbracket} to functions of the type $\mathsf{H}(\bz,\Sigma)$. This process yields the direct sum bracket 
\begin{align}\label{nonlocalbracket2}
\{f,g\}=&\{f,g\}_{\bz}+2\operatorname{Tr}\!
\left(\!\Sigma\left(\frac{\delta {f}}{\delta \Sigma}\Bbb{J}\frac{\delta {g}}{\delta \Sigma}-\frac{\delta {g}}{\delta \Sigma}\Bbb{J}\frac{\delta {f}}{\delta \Sigma}\right)\!\right)
.
\end{align}
{\color{black}This Poisson bracket produces the following equations for a total energy of the form $h=h(\bz,\Sigma)$:
\begin{equation}
\dot{\bz} = \{\bz,h\}_c
\,,\qquad\ 
\dot{\Sigma} = 2 \left( \Bbb{J}\frac{\delta h}{\delta \Sigma}\Sigma + \Sigma\frac{\delta h}{\delta \Sigma}\Bbb{J} \right).
\label{varianceformulation}
\end{equation}
For example, in the particular case when the total energy $\langle H\rangle =\int \!G(\bzeta) H(\bzeta)\,\de\zeta$ is approximated by using \eqref{Taylor},
these equations recover the dynamics (12) and (13) in \cite{GrSch}, suitably specialized to Hermitian quantum mechanics. 
The explicit comparison of the equations \eqref{varianceformulation} with those obtained by Heller for Gaussian wavepackets (see \cite{Heller1,Heller2} and subsequent papers by Heller on the same topic) requires expressing the covariance matrix as a function on the Siegel upper half-space \cite{OhLe,Ohsawa}. The corresponding geometric description is the subject of ongoing work. 

Equations \eqref{varianceformulation} can be directly applied to modify certain moment models that have} previously appeared in the  chemical physics literature \cite{Prezhdo2,PrezhdoPereverzev,PrezhdoPereverzev2}. This class of models suffers from lack of energy conservation in the general case \cite{PrezhdoPereverzev2}, with possible consequent drawbacks on the time evolution properties. In references \cite{Prezhdo2,PrezhdoPereverzev,PrezhdoPereverzev2} and related papers on the same topic, a class of moment models was developed  by adopting a Gaussian moment closure on the equations of motion for the expectation values $\langle \bZeta \rangle$ and $\langle\bZeta\bZeta\rangle$. More particularly,  Gaussian closures of the type $\langle\bZeta\bZeta\bZeta\rangle\simeq3\langle\bZeta\bZeta\rangle\langle\bZeta\rangle-2\langle\bZeta\rangle^3$ (and similarly for higher order moments) are used in the equations 
\[
i\hbar\frac{\de}{\de t}\langle{\bZeta}\rangle =\big\langle \big[ {\bZeta},\widehat{H} \big] \big\rangle
\,,\qquad\qquad
i\hbar\frac{\de}{\de t}\langle{\bZeta\bZeta}\rangle =\big\langle \big[ {\bZeta\bZeta},\widehat{H} \big] \big\rangle\,.
\]
The moment closures are performed after replacing the Hamiltonian operator $\widehat{H}$ by its Taylor expansion around the expectation value $\bz=\langle\bZeta\rangle$ (similarly to \eqref{Taylor}). When this Taylor expansion is truncated to third order, one recovers the Gaussian moment closure previously introduced in \cite{TuSn}. However, in the general case of a higher-order expansion, adopting the closure directly in the Gaussian moment equations may break energy conservation  \cite{PrezhdoPereverzev2}. Indeed, while odd-order expansions of $\widehat{H}$ preserve  energy conservation, this does not hold for truncations of even order greater than 2. 
{\color{black}More particularly, as noticed in \cite{PrezhdoPereverzev2}, in order to conserve the total energy, the Taylor expansion used in the equations should be terminated after an odd derivative.} 
 The framework presented in this Section provides a solution to this problem. Indeed, once the closure has been performed in the expression of the total energy $\mathsf{H}=\langle\widehat{H}\rangle$, the Hamiltonian moment equations \eqref{Jack1}-\eqref{Jack2} are uniquely determined.  As a consequence of the Poisson bracket structure \eqref{EQ:QmomPB}, these equations generally differ from those in \cite{PrezhdoPereverzev2} by {\color{black}exactly  the odd derivatives} that ensure conservation of both the total energy and the determinant of the covariance matrix. More particularly, if one expands the Hamiltonian $\widehat{H}=\widehat{P}^2/2+V(\widehat{Q})$ up to fourth order around the expectation values $\langle{\bZeta}\rangle$, the second parenthesis in equation \eqref{Jack2} produces conservative terms consisting of the fifth-order derivatives in the following equations of motion (in standard expectation value notation):
\begin{align*}
\frac{\de}{\de t} \mQ =&\, \mP\,,
\\
\frac{\de}{\de t} \mP =&\, -V^{(1)}\!(\mQ) - \frac{1}{2} V^{(3)}\!(\mQ)\,(\mQs-\mQ^2) - \frac{1}{8} V^{(5)}\!(\mQ)\,(\mQs-\mQ^{2})^{2}\,,
\\
\frac{\de}{\de t}\mQs =&\, 2 \mQP_{s}\,,
\end{align*}
\begin{align*}
\frac{\de}{\de t}\mPs =&\, -2 V^{(1)}\!(\mQ)\,\mP - 2 V^{(2)}\!(\mQ)\,(\mQP_{s}-\mQ\mP) - V^{(3)}\!(\mQ)\, \mP (\mQs-\mQ^{2}) +
\\
&\,- V^{(4)}\!(\mQ)\, (\mQP_{s}-\mQ\mP)(\mQs-\mQ^{2}) - \frac{1}{4} V^{(5)}\!(\mQ)\, \mP (\mQs-\mQ^{2})^{2}\,,
\\
\frac{\de}{\de t}\mQP_{s} =&\, \mPs - V^{(1)}\!(\mQ)\,\mQ - V^{(2)}\!(\mQ)\,(\mQs-\mQ^{2}) - \frac{1}{2} V^{(3)}\!(\mQ)\,\mQ (\mQs-\mQ^{2})   +
\\
&\,- \frac{1}{2} V^{(4)}\!(\mQ) (\mQs-\mQ^{2})^{2} - \frac{1}{8} V^{(5)}(\mQ) \mQ (\mQs-\mQ^{2})^{2}\,,
\end{align*}
where $\mQP_{s}$ denotes the expectation of the symmetrized product $(\widehat{Q}\widehat{P}+\widehat{Q}\widehat{P})/2$. Dropping the fifth-order derivatives returns the non-conservative equations  (15)--(19) in \cite{PrezhdoPereverzev2}. Therefore, for exact polynomial potentials $V(\widehat{Q})$ of fourth degree, energy conservation is not an issue because $V^{(5)}(\mQ)\equiv0$ and the two models coincide. However, more general cases such as Morse-type potentials require extra care in dealing with the Gaussian closure. The implications of the energy-conserving terms in concrete physical problems will be the subject of future work.


\section{Conclusions and perspectives}
Based on the Hamiltonian Poisson bracket approach, this paper has unfolded the geometric properties of Ehrenfest's expectation value dynamics. {\color{black} More particularly, the search for the Hamiltonian structure of Ehrenfest's theorem has produced a new classical-quantum Poisson structure that incorporates classical and quantum dynamics as special cases. The corresponding equations are Lie-Poisson for the Ehrenfest group $\mathcal{H}(\Bbb{R}^{2n})\circledS U(\mathscr{H})$. This result was achieved by exploiting the momentum map property underlying Ehrenfest expectation values. Later, the Ehrenfest theorem was rewritten upon expressing quantum dynamics in the phase-space frame co-moving with the expectation values. This result extends previous work in quantum cosmology \cite{Bojowald,BoSh} by avoiding moment truncation problems.} In the last part, the Poisson structures underling Ehrenfest theorem were restricted to consider Gaussian moment dynamics, as it arises from Gaussian Wigner functions on phase-space. Again, Gaussian moments enjoy a momentum map structure that confers them a Lie-Poisson bracket for the Jacobi group $\operatorname{Sp}(\Bbb{R}^{2n}) \circledS\mathcal{H}(\Bbb{R}^{2n})$. This bracket structure enables providing  energy-conserving variants of previous Gaussian moment models \cite{Prezhdo2,PrezhdoPereverzev} that were generally lacking conservation of energy  \cite{PrezhdoPereverzev2}. 

This paper has shown that the use of Wigner functions and the properties of the Moyal bracket are particularly advantageous for studying expectation values. Then, combining Poisson brackets with momentum map structures unfolds the geometry underlying quantum dynamics. For example, momentum map structures may also appear in quantum hydrodynamics, where local averages (e.g. $\int\! W(\boldsymbol{q},\boldsymbol{p})\,\de^3\boldsymbol{p}$ and $\int\!\boldsymbol{p}\, W(\boldsymbol{q},\boldsymbol{p})\,\de^3\boldsymbol{p}$) are considered. In addition, although the present treatment did not consider spin effects, these can be retained by including the expectation $\langle \boldsymbol{\widehat{S}}\rangle$ of the spin operator $\boldsymbol{\widehat{S}}$ in the treatment of the classical-quantum bracket \eqref{hybridPB}. Another open question concerns the ladder operator formulation of these results. This can be particularly advantageous for the study of coherent squeezed states in quantum optics and is currently under development by the authors. Future research directions will also involve the Heisenberg picture and the corresponding variational framework.

\medskip
\paragraph{Acknowledgments.} The authors are indebted to Tomoki Ohsawa, Alessandro Torrielli and the anonymous referees for providing extensive and valuable feedback on these results. In addition, the authors are grateful to Dorje Brody, Fran\c{c}ois Gay-Balmaz, Darryl Holm, David Meier, Juan-Pablo Ortega and Paul Skerritt for several discussions on this and related topics. This work was partially carried out at the Bernoulli Center of the Swiss Federal Institute of Technology in Lausanne: C.T. acknowledges hospitality during the program ``Geometric Mechanics, Variational and Stochastic Methods''. Financial support by the Leverhulme Trust Research Project Grant 2014-112, the London Mathematical Society Grant No. 31320 (Applied Geometric Mechanics Network), and the  EPSRC Grant No. EP/K503186/1  is also acknowledged.

\rem{ 
\paragraph{Data accessibility statement.} This work contains no simulation or experimental data.

\paragraph{Ethics statement.}Ethical considerations are inapplicable to the work presented here.

\paragraph{Competing interests.}
The authors have no competing interests to declare.

\paragraph{Authors' contributions.}
E.B.L. focused mainly on Sections 4 and 5, while C.T. worked on the remaining parts of the paper.

\paragraph{Funding statement.} Financial support by the Leverhulme Trust Research Project Grant 2014-112, the London Mathematical Society Grant No. 31320 (Applied Geometric Mechanics Network), and the  EPSRC Grant No. EP/K503186/1  is also acknowledged.

} 

\end{document}